\documentclass[aps,prl,twocolumn,superscriptaddress]{revtex4-2}
\usepackage{graphicx}
\usepackage{amsmath}
\usepackage{amssymb}
\usepackage{colordvi}
\usepackage{mathrsfs}
\usepackage{bm}
\usepackage{verbatim}
\usepackage{dcolumn}
\usepackage{epsfig}
\usepackage{subfigure}
\usepackage{makecell}
\usepackage[colorlinks,linkcolor=blue,anchorcolor=blue,citecolor=blue,urlcolor=blue]{hyperref}
\usepackage{float}

\begin{document}

\title{Quantum-metric-nematicity induced Kerr-like polarization rotation without time-reversal symmetry breaking}
\author{Wenhao Liang}
\thanks{Contact author: wenhaoliang@ust.hk}
\affiliation{Department of Physics, Hong Kong University of Science and Technology, Clear Water Bay, Hong Kong, China}

\author{Akito Daido}
\affiliation{Department of Physics, Graduate School of Science, Kyoto University, Kyoto 606-8502, Japan}
\affiliation{Department of Physics, Hong Kong University of Science and Technology, Clear Water Bay, Hong Kong, China}

\author{K. T. Law}
\affiliation{Department of Physics, Hong Kong University of Science and Technology, Clear Water Bay, Hong Kong, China}

\date{\today{}}

\begin{abstract}
The magneto-optic Kerr effect (MOKE), which describes the rotation and ellipticity of linearly polarized light upon reflection, is conventionally associated with time-reversal symmetry breaking.
Here, we theoretically demonstrate that a Kerr-like polarization rotation can emerge even in nonmagnetic systems with time-reversal symmetry, owing to the nontrivial quantum metric of electronic bands. 
We show that the nematicity of the quantum metric, which captures the anisotropy of the quantum metric tensor due to the breaking of $n$-fold (with $n \ge 3$) rotational symmetry, gives rise to an incident-polarization-dependent reflected-polarization rotation. 
Notably, this mechanism requires neither magnetic order nor spin-orbit coupling, which are conventionally considered essential for MOKE. 
We illustrate the effect using a minimal tight-binding model and a model for strained MoS$_2$. 
This work reveals a quantum-geometric origin of the polarization rotation effects beyond conventional MOKE and suggests a new experimental approach to detect quantum metric nematicity.
\end{abstract}

\maketitle
\emph{Introduction.---}
In 1877, Kerr observed that when linearly polarized light is reflected from a magnet, it transforms into elliptically polarized light with a rotated polarization axis \cite{Kerr1877}. Ever since, magneto-optic Kerr effect (MOKE) has been studied for over a century in various systems breaking time-reversal ($\mathcal{T}$) symmetry through spontaneous magnetization or external magnetic fields~\cite{ Weinberger2008,Feil1987, Pershan1967, Qiu2000, Budker2002}. Such systems include ferromagnets \cite{Argyres1955, Erskine1973, Huebner1989, Hamrle2002, Kim2020, Guo1995, Guo1994, Kim1999, Stroppa2008, Ravindran1999, Huang2017, Lyalin2023, Rathgen2005}, antiferromagnets \cite{Sivadas2016, Feng2015, Li2025, Ahn2022, Higo2018, Feng2020, Iguchi2023, Li2025b,Wu2020}, ferrimagnets \cite{Fleischer2018, Johnson2023}, kagome superconductors \cite{Xu2022}, and other topological materials \cite{Tse2011, Li2024, Kato2023, Habibi2022}. 
In modern theories of MOKE, the rotation of the polarization is attributed to $\mathcal{T}$ symmetry breaking, which generates antisymmetric off-diagonal components in the optical conductivity tensor through the Berry curvature~\cite{Feng2015,Kato2023,Wu2020}. 
Additionally, the spin-orbit coupling (SOC) has been considered essential, as it couples magnetization to electrons' orbital motions~\cite{Ebert1996, Sivadas2016, Feng2015}.

In recent studies of optical responses as well as transport and various other phenomena, the concept of quantum geometry has attracted significant attention~\cite{Toermae2023,Chen2022,Resta2011,Verma2026}.
While the Berry curvature has been extensively studied over the past few decades~\cite{Chang2023, Liang2025, Hasan2010,Sodemann2015, Liang2023, Xiao2010,Bac2025, Liang2023a}, the quantum metric, which measures the distance between neighboring Bloch states in Brillouin zone, has attracted broader attention only recently.
The quantum metric has already been shown to play an important role in nonlinear optical responses \cite{Holder2020,Ahn2020,Ahn2021,Hsu2023,Ghosh2024},
nonlinear transport \cite{Gao2023,Wang2023},
electrical properties \cite{Komissarov2024, Verma2025}, and flat-band superconductivity \cite{Toermae2018,Huhtinen2022,Toermae2022, Chen2024, Penttilae2025,Guo2025}. 
However, despite  significant interest and research efforts,
it still remains challenging to develop methods to directly measure quantum metric quantities  \cite{Toermae2023,Balut2025,Verma2025}. 

\begin{figure}
    \centering
    \includegraphics[width=\linewidth]{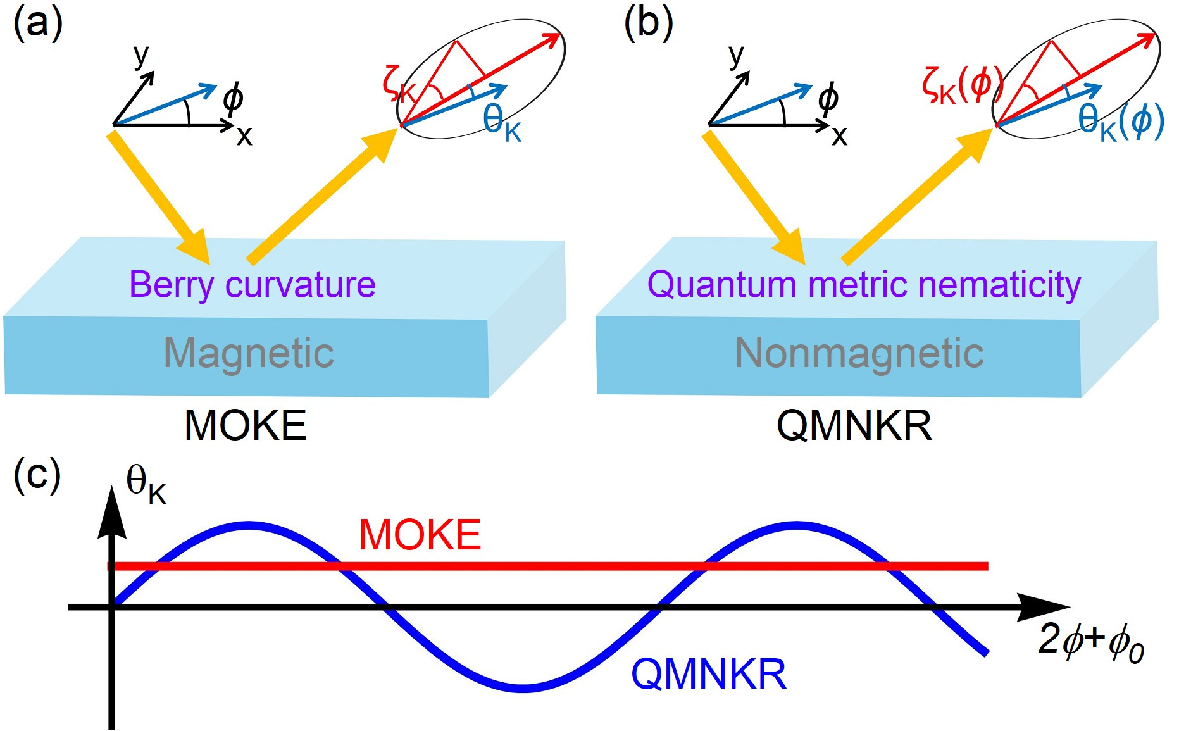}
    \caption{Schematic figure of the (a) MOKE and (b) QMNKR. The incident light is affected by (a) the Berry curvature and (b) the quantum-metric nematicity and reflected as an elliptically-polarized light, and experiences a polarization rotation $\theta_K$ (red arrow) and gains an ellipticity $\zeta_K$, which depends on the incident polarization angle $\phi$ (blue arrow) only for the QMNKR. (c) The MOKE is independent of $\phi$, whereas the QMNKR exhibits a twofold angular dependence, where $\phi_0$ is a material-dependent parameter.
    }
    \label{fig:QMNKR}
\end{figure}

In this Letter, we point out that in contrast to previous belief, $\mathcal{T}$-symmetry breaking and SOC are not necessary for Kerr-like rotation.
We elucidate that the Kerr-like polarization rotation effects can emerge when the system has quantum metric with nematicity, i.e., the breaking of rotational symmetry higher than twofold. And we name such unconventional Kerr rotation in nonmagnetic systems without $\mathcal{T}$ symmetry breaking as the quantum-metric-nematicity induced Kerr-like rotation (QMNKR). Furthermore, we show that the polarization rotation angle $\theta_K$ can be written as  $\theta_K = \theta_{\text{MOKE}} + \theta_{QM} \text{sin}(2\phi+\phi_0)$ in weakly anisotropic systems, where the $\theta_{\text{MOKE}}$ is the conventional MOKE contribution due to $\mathcal{T}$ symmetry breaking, and $\theta_{QM} \text{sin}(2\phi+\phi_0)$ is the quantum metric contribution which depends on the incident polarization angle $\phi$ which can appear in the presence of $\mathcal{T}$ symmetry [Fig.~\ref{fig:QMNKR}].
Interestingly, a polarization-rotation signal with the same leading twofold angular dependence has been reported in nematic kagome superconductors~\cite{Xu2022}, and the angle-dependent contribution was analyzed phenomenologically in terms of birefringence associated with rotational-symmetry breaking. Here we show that, within a two-dimensional optical-conductivity framework, quantum-metric nematicity provides an alternative microscopic mechanism for a similar angle-dependent response. 

We demonstrate the presence of the $\theta_{QM}$ term by using a minimal model and strained MoS$_2$ that quantum-metric nematicity is the essential ingredient of such Kerr-like rotation without $\mathcal{T}$ symmetry breaking. The observation of the $\theta_{QM}$ provides evidence for finite quantum metric nematicity.

\emph{Optical conductivity and quantum geometry.---} In this section, we introduce the connection between optical conductivity and quantum geometric tensor, which incorporates both the Berry curvature and the quantum metric effects. 
To illustrate the concept, we focus on two-dimensional insulator at zero temperature in this Letter. In this case, the intraband terms can be neglected, and the frequency dependent optical conductivity can be expressed as ~\cite{SM}
\begin{align}\label{eq_sigma}
	\sigma_{\alpha\beta}(\omega) &= \frac{e^2}{\hbar}\sum_{m\neq n} \int \frac{d \bm{k}}{(2 \pi)^2} f_{m}(1-f_{n})\frac{\varepsilon_{m n }^2}{\varepsilon_{m n }^2-(\hbar \omega+i \eta)^2} \notag \\
	&\quad \times \left( -\frac{\varepsilon_{m n }}{\hbar\omega} \Omega^{mn}_{\alpha\beta} + 2i  g^{mn}_{\alpha\beta} \right).
\end{align}
Here, $\Omega^{mn}_{\alpha\beta}=i(A^{mn}_\alpha A^{nm}_\beta-A^{mn}_\beta A^{nm}_\alpha)$ 
 and $ g^{mn}_{\alpha\beta}=\frac{1}{2}(A^{mn}_\alpha A^{nm}_\beta+A^{mn}_\beta A^{nm}_\alpha)$ denote the band-resolved Berry curvature and quantum metric, respectively. The Berry connection are denoted by $A^{mn}_\alpha=i\left\langle u_m|\partial_\alpha u_n\right\rangle $, $f_{n}=0,1$ is the Fermi distribution function, 
 $\varepsilon_{mn}=\varepsilon_{m}-\varepsilon_{n}$ with $\varepsilon_{n}$ being the energy dispersion for band $n$, and $\eta $ is the smearing parameter in units of energy.

From a symmetry perspective, the Berry curvature vanishes on average in $\mathcal{T}$-symmetric systems. In this case, the optical conductivity is  entirely determined by the quantum metric contribution and is a symmetric tensor.
The optical conductivity tensor is further constrained by spatial symmetries: For instance, in systems with $n$-fold rotational symmetry $C_{nz} (n\geq 3)$, the relations $\sigma_{xx}=\sigma_{yy}$ and $\sigma_{xy}=-\sigma_{yx}$ are satisfied.
This indicates that the rotational symmetry breaking, or nematicity, of the quantum metric plays an important role in nontrivial optical properties.

\emph{Rotation angle and ellipticity}.---To clarify the quantum-metric effect on reflection, we derive a general formula for the polarization rotation and ellipticity in general two-dimensional materials.
Consider normally incident light propagating along the $z$-direction and linearly polarized in the $xy$ plane at an angle $\phi$ relative to the $x$-axis,
$\bm E_{\rm i}(z,t)=E_{\rm i} e^{i(qz-\omega t)}\hat{\bm e}_{\rm i}$,
where $\hat{\bm e}_{\rm i}=(\cos \phi,\sin \phi)$.
By solving Maxwell's equations and applying boundary conditions at the interface $z=0$, we obtain the reflection tensor $r_{\alpha\beta}$ by~\cite{SM} 
\begin{align}\label{eq_r}
	r_{\alpha\beta} =\frac{-Z_0(2\sigma_{\alpha\beta}+Z_0\delta_{\alpha\beta} \det \sigma)}{4+2Z_0\operatorname{tr}\sigma+Z_0^2 \det
	\sigma},
\end{align}
where $Z_0=\sqrt{\frac{\mu_0}{\epsilon_0}}$ is the impedance of the free space, $\epsilon_0$ is the dielectric constant and $\mu_0$ the permeability in the vacuum. Here, $\sigma_{\alpha\beta}=\sigma_{\alpha\beta}(\omega)$ is the optical conductivity tensor.
The reflected electric field can be expressed as $\bm{E}_{\rm r}=rE_{\rm i}\hat{\bm{e}}_{\rm i}$ with Eq.~\eqref{eq_r}, which generally has components both parallel and perpendicular to the incident polarization $\hat{\bm{e}}_{\rm i}$ with phase difference~(see Supplementary Material~\cite{SM}).
This means that the reflected electric field
$\bm E_{\rm r}$ becomes elliptically polarized, in contrast to the linearly polarized incident light $\bm E_{\rm i}$.

Having identified the reflected field is elliptically polarized, the complex Kerr angle can be obtained through geometric analysis \cite{SM}:
\begin{align}\label{eq_angle1}
	\theta_{\rm K} + i\zeta_{\rm K} =
    \frac{-\sigma_{\rm H}+\bar{\sigma}_{xy}\cos 2\phi - \bar{\sigma}_{x^2-y^2}\sin2\phi}{\bar{\sigma}_{x^2-y^2}\cos2\phi + \bar{\sigma}_{xy}\sin2\phi + \frac{\operatorname{tr}\sigma + Z_0\det\sigma}{2}}.
\end{align}
Here, $\theta_{\rm K}$ represents the rotation angle of the major axis of the reflected light relative to the polarization direction of the incident light, while $\zeta_{\rm K}$ denotes the ratio of the major to minor axis of the polarization ellipse. 
We can see that the optical conductivity enters through the Hall component $\sigma_{\rm H}=(\sigma_{xy}-\sigma_{yx})/2$ and longitudinal and transverse anisotropic components $\bar{\sigma}_{x^2-y^2}=(\sigma_{xx}-\sigma_{yy})/2$ and $\bar{\sigma}_{xy}=(\sigma_{xy}+\sigma_{yx})/2$, respectively.

Equation~\eqref{eq_angle1} shows that the reflected-polarization rotation generally acquires a twofold angular dependence when the optical conductivity has anisotropic components.
When the anisotropy is weak, this dependence is captured by
$\theta_K = \theta_{\text{MOKE}} + \theta_{QM} \text{sin}(2\phi+\phi_0)$.  
Here, $\phi$ is the polarization angle of the incident light and $\phi_0$ is a material dependent parameter. 
This has the same leading angular dependence as in the phenomenological decomposition used in a recent experiment on kagome superconductors 
below $T_{\rm CDW}$~\cite{Xu2022}, suggesting that Eq.~\eqref{eq_angle1} provides, within the present two-dimensional optical-conductivity framework, a microscopic realization of the angle-dependent term with quantum-metric origin.

It should be noted that the expression Eq.~\eqref{eq_angle1} holds regardless of whether the system has $\mathcal{T}$ symmetry. In systems with $C_{nz} (n\geq 3)$ symmetry, the anisotropic components $\bar{\sigma}_{x^2-y^2}$ and $\bar{\sigma}_{xy}$ are prohibited, and 
$\theta_{\rm K}$ is solely determined by the Hall component and $\phi$ independent: This corresponds to the conventional MOKE and the $\theta_{QM}$ term is zero.
On the other hand, even in systems with $\mathcal{T}$ symmetry and $\sigma_{\rm H}=0$, we can obtain the Kerr-like rotation as long as $\bar{\sigma}_{x^2-y^2}$ or $\bar{\sigma}_{xy}$ are finite when $C_{nz} (n\geq 3)$ symmetry is broken.
As can be seen from Eq.~\eqref{eq_sigma}, $\bar{\sigma}_{x^2-y^2}$ or $\bar{\sigma}_{xy}$ can be nonzero when the quantum-metric nematicity 
\begin{align}
    \bar{g}^{mn}_{{x^2-y^2}}(\bm{k})&=\frac{1}{2}\left( g_{xx}^{mn}(\bm{k})-g_{yy}^{mn}(\bm{k})\right) ,\notag\\
   \bar{g}^{mn}_{{xy}}(\bm{k})&=\frac{1}{2}\left( g_{xy}^{mn}(\bm{k})+g_{yx}^{mn}(\bm{k})\right) =g_{xy}^{mn}(\bm{k}),
\end{align}
are nonzero and capture their overall behavior across the Brillouin zone.
Thus, the observation of Kerr-like rotation in $\mathcal{T}$-symmetric systems, the $\theta_{QM}$ term, is the manifestation of the quantum-metric nematicity.

\begin{figure*}
	\includegraphics[width=17cm,angle=0]{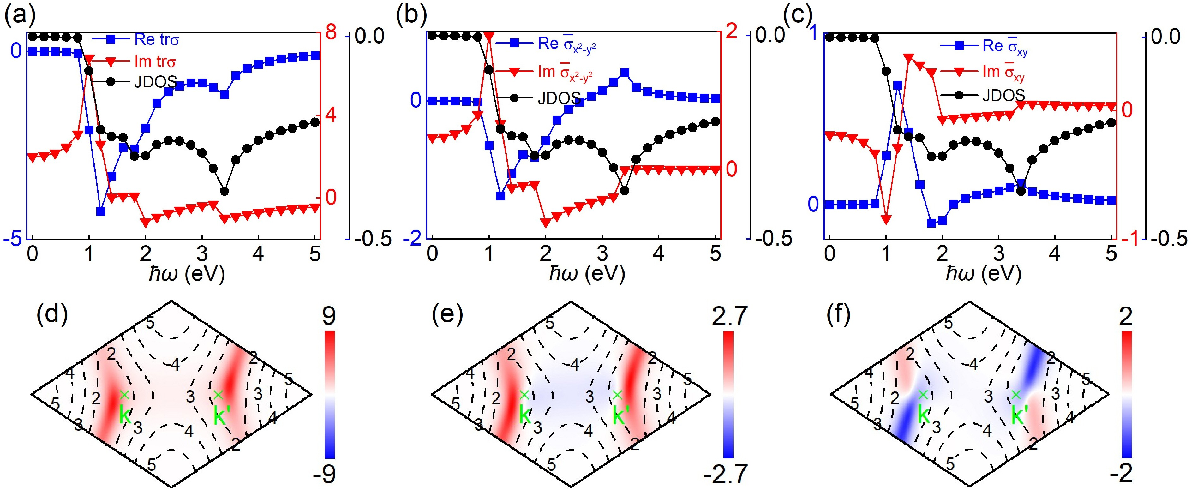}
	\caption{The optical conductivity (in units of $e^2/h$) and quantum metric nematicity in the nonmagnetic minimal model. The optical conductivity (a) $\text{tr} \sigma$, (b) $\bar{\sigma}_{x^2-y^2}$, (c) $\bar{\sigma}_{xy}$ and corresponding JDOS with the smearing parameter $\eta=0.01$ eV. The distribution of quantum metric quantities (d) $\text{tr} g$, (e) $\bar{g}_{x^2-y^2}$, and (f) $\bar{g}_{xy}$ in the first Brillouin zone, with $``\times"$ denoting the position of two valleys K and $\text{K}^\prime$. The dashed curve shows the equal energy contour for the optical transition with different photon energy. }
	\label{graphene1}
\end{figure*}

\emph{QMNKR in minimal model.---}We illustrate the QMNKR with a minimal tight-binding model on a honeycomb lattice  
described by the Hamiltonian
\begin{align}\label{eq_H}
	H = - \sum_{\langle i j \rangle} t_{ij}c_{i}^{\dagger} c_{j} + \sum_{i} c_{i}^{\dagger} \Delta_i c_{i},
\end{align}
with $i$, $j={\rm A,B}$ being the sublattice degrees of freedom. 
The second term is the sublattice potential $\Delta_A=-\Delta_B=\Delta$, which introduces a mass term to the Dirac points to open the band gap.
The first term is the nearest-neighbor hopping with different amplitudes $t_1$, $t_2$, and $t_3$ for three different directions.
This model realizes massive Dirac models with pseudo-gauge fields, and may correspond to graphene coupled to hBN~\cite{Giovannetti2007} with applied strain~\cite{Choi2010}. 
We set $t_1=1$ eV as energy unit, 
$t_2=1.2 t_1$, $t_3=0.6 t_1$, and $\Delta=0.5 t_1$.

The model preserves $\mathcal{T}$ symmetry and breaks $C_{nz} (n\ge3)$ symmetry. Therefore, the system is allowed to show the QMNKR, while the MOKE is prohibited.
To see this, we first show in
Figs.~\ref{graphene1}(a)–(c) the optical conductivity tensor computed from Eq.~\eqref{eq_sigma}. 
Not only the isotropic component $\mathrm{tr}\sigma$ but also the anisotropic components $\bar{\sigma}_{x^2-y^2}$ and $\bar{\sigma}_{xy}$ are finite, reflecting the $C_{nz} (n\ge3)$ symmetry breaking.
The behavior of the isotropic component $\mathrm{tr}\sigma$ can qualitatively be understood by the joint-density of states (JDOS),
\begin{align}
\text{JDOS} = \sum_{m \neq n} \int \frac{d\bm k}{(2\pi)^2} f_{mn}\delta(\varepsilon_{mn} - \hbar \omega).
\end{align}
Indeed, the overall spectrum of $\mathrm{tr}\,\sigma(\omega)$ (blue) is well reproduced by that of JDOS (black) [Fig.~\ref{graphene1}(a)].
For example, when the photon energy $\hbar\omega$ is smaller than the band gap $\approx1$ eV, 
JDOS is absent, and therefore $\mathrm{Re}\,\sigma_{\alpha\beta}(\omega)$ vanishes.
We can also see that the peak of $\mathrm{Re}\,\mathrm{tr}\,\sigma(\omega)$ at $\hbar\omega=3.4$ eV follows from that of the JDOS. 

On the other hand, the anisotropic components of the optical conductivity $\bar{\sigma}_{x^2-y^2}$ and $\bar{\sigma}_{xy}$ have spectrum much different from JDOS.
In contrast to JDOS, which has no sign change,
$\mathrm{Re}\,\bar{\sigma}_{x^2-y^2}$ and $\mathrm{Re}\,\bar{\sigma}_{xy}$ show sign reversals at $\hbar\omega\approx2.5$ eV and $1.7$ eV in Figs.~\ref{graphene1}(b) and (c), respectively.
These results clearly illustrate the importance of the quantum metric nematicity in understanding anisotropic components of the optical conductivity.

To clarify the role of quantum metric, we plot its distribution over the Brilloiuin zone as shown in Fig.~\ref{graphene1}(d)-(f).
In our model, the band gap opens slightly away from the K and $\text{K}^\prime$ points due to the strain (the pseudo-gauge field effects), and gives rise to the quantum-metric hot spots nearby.
The quantum-metric nematicity $\bar{g}_{x^2-y^2}(\bm{k})$ and $\bar{g}_{xy}(\bm{k})$ are shown in Figs.~\ref{graphene1}(e) and (f).
They show both positive and negative regions in Brillouin zone, in contrast to the isotropic component $\mathrm{tr}\,g(\bm{k})\ge0$ [Fig.~\ref{graphene1}(d)].

Such a separation of positive and negative quantum-metric nematicity in Brillouin zone accompanies that in energy, and this is essential to understand the behavior of $\bar{\sigma}_{x^2-y^2}$ and $\bar{\sigma}_{xy}$.
Indeed, the quantum-metric nematicity averaged over the energy contour $\epsilon_{mn}(\bm{k})=\hbar\omega$
determines the optical conductivity up to a prefactor and JDOS, as understood from the formula~\cite{SM}
\begin{align}
    \mathrm{Re}\,\bar{\sigma}_{x^2-y^2, xy}(\omega) 
	=\pi\omega e^2[\text{JDOS}]\langle{\bar{g}_{x^2-y^2, xy}}\rangle_\omega,
\end{align}
with the averaged quantum-metric nematicity on the resonant energy contour
\begin{align}
    	\langle{\bar{g}_{x^2-y^2, xy}}\rangle_\omega\equiv \frac{\sum_{\substack{m\in occ, n\notin occ}} \int d \bm{k} \delta(\varepsilon_{m n }+\hbar\omega) \bar{g}^{mn}_{x^2-y^2, xy}}{\sum_{\substack{m\in occ, n\notin occ}} \int d \bm{k} \delta(\varepsilon_{m n }+\hbar\omega)}.
\end{align}
In Fig.~\ref{graphene1}(e) (Fig.~\ref{graphene1}(f)),the quantum-metric nematicity changes sign near $\hbar\omega =2.5$ ($\hbar\omega =1.7$ ) eV as can be read out from the energy contours shown by dashed lines. This explains the sign reversals of $\mathrm{Re}\,\bar{\sigma}_{x^2-y^2}$ and 
$\mathrm{Re}\,\bar{\sigma}_{xy}$ in Fig.~\ref{graphene1} (b) and (c).

Next, we discuss the imaginary part of the optical conductivity [red curves in Fig.~\ref{graphene1}(a)-(c)].
While the imaginary part has a complicated spectrum due to the energy factor in Eq.~\eqref{eq_sigma}, the quantum-metric nematicity is still required for the anisotropic components $\mathrm{Im}\,\bar{\sigma}_{x^2-y^2}$ and $\mathrm{Im}\,\bar{\sigma}_{xy}$ to be finite, as ensured by symmetry.
A particularly simple relation between them is obtained when the frequency is much smaller than the band gap, where the optical conductivity is proportional to the quantum-metric nematicity integrated over the Brillouin zone, 
$\bar{\sigma}_{x^2-y^2,xy}(\omega\to0)=(2ie^2/\hbar)\sum_{m\neq n}\int\frac{d\bm{k}}{(2\pi)^2}\bar{g}^{mn}_{x^2-y^2,xy}$. 
This corresponds to the quantum weight~\cite{Onishi2024, Onishi2025} and provides the finite offset at $\omega=0$ in Fig.~\ref{graphene1}(b) and (c).
Moreover, it's worth noting that the quantum weight is also related to $\operatorname{Re}\sigma(\omega)$ through the optical sum rule{(see Supplementary
Material~\cite{SM})}, and thus QMNKR proves a way to measure the quantum weight.

\begin{figure}
	\includegraphics[width=8.5cm,angle=0]{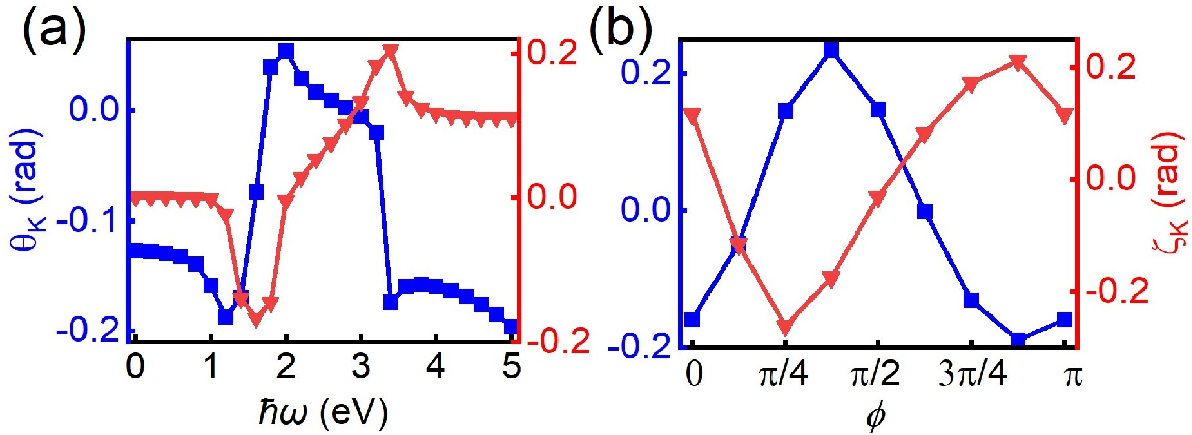}
	\caption{QMNKR in nonmagnetic minimal model. The rotation angle $\theta_K$ and ellipticity $\zeta_K$ (a) versus photon energy $\hbar \omega$ ($\phi=0$), and (b) versus polarization angle $\phi$ ($\hbar \omega=4$ eV).}
	\label{graphene2}
\end{figure}

To quantify the QMNKR, we discuss the complex Kerr angle $\theta_{\rm K}+i\zeta_{\rm K}$. 
Figure~\ref{graphene2}(a) shows the $\omega$ dependence of $\theta_{\rm K}$ (blue) and $\zeta_{\rm K}$ (red) for a fixed polarization direction $\phi=0$, where $\bar{\sigma}_{xy}$ is essential according to Eq.~\eqref{eq_angle1}.
When the photon energy is below the band gap, the ellipticity $\zeta_K$ vanishes due to the absence of absorption, while the rotation angle $\theta_K$ remains finite. 
As the photon energy increases, the interplay between the quantum metric and the energy factor leads to the evolution of the complex Kerr angle, accompanied by several pronounced peaks. 
Moreover, increasing the mass term $\Delta$ to enlarge the band gap roughly reduces the peak values of Kerr angles~\cite{SM}, as the quantum metric generally decreases with increasing band gap, which also demonstrate the key role of the quantum metric.
In Fig.~\ref{graphene2}(b), we fix the photon energy at $\hbar\omega=4$ eV and examine the dependence of the complex Kerr angle on the incident polarization direction $0\le\phi\le\pi$.
The QMNKR is not only obtained at $\phi=0$ but also at $\phi=\pi/4$, reflecting the fact that both of the quantum-metric nematicity $\bar{g}_{xy}$ and $\bar{g}_{x^2-y^2}$ are finite in our model.
Importantly, the $\theta_K$ exhibits a $\text{sin}(2\phi+\phi_0)$-like behavior, which is different from conventional $\phi$-independent MOKE.

\emph{QMNKR in strained MoS$_2$.---}We established that the QMNKR emerges owing to the quantum-metric nematicity by using a minimal model, where the system preserves $\mathcal{T}$ symmetry and lacks SOC and thus conventional MOKE is prohibited. 
To demonstrate its feasibility in a realistic material platform, we examine the QMNKR in monolayer MoS$_2$ with strain.
We adopt a model parametrized from first-principles calculations~\cite{SM}, and the strain effect is taken into account by phenomenologically modifying the hopping parameters as in the previous model, to ensure the presence of the quantum-metric nematicity.
The applied strain is estimated to be about $2\%$~\cite{SM} and is well achievable in experiments~\cite{Manzeli2017}.

\begin{figure}
	\includegraphics[width=8.5cm,angle=0]{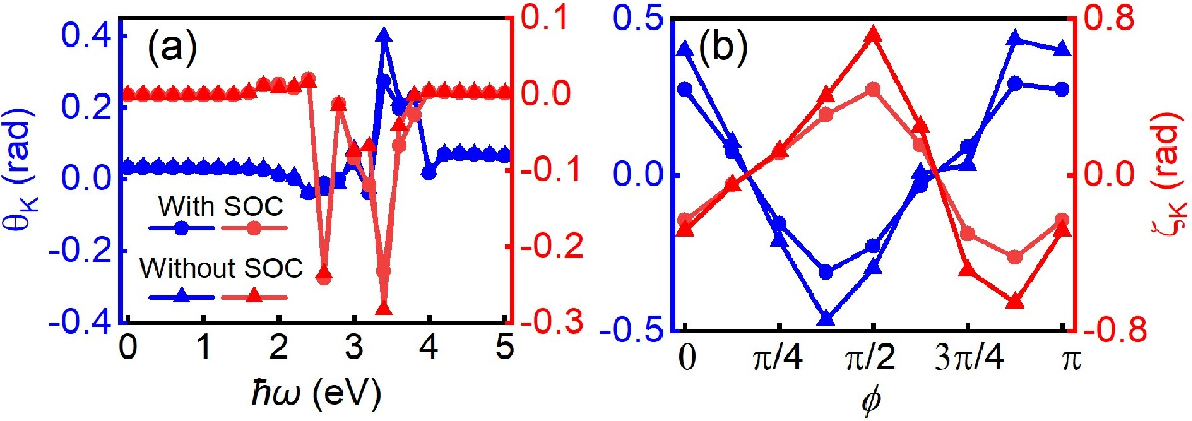}
	\caption{QMNKR in strained MoS$_2$ with SOC turned on and turned off respectively. The $\theta_K$ and $\zeta_K$ (a) versus photon energy $\hbar \omega$ ($\phi=0$), and (b) versus polarization angle $\phi$ ($\hbar \omega=3.4 eV$).}
	\label{angles_TMD}
\end{figure}

The results of the QMNKR are shown in Fig.~\ref{angles_TMD},which are evaluated with SOC turned on and off, respectively.
The photon-energy dependence of the complex Kerr angle for the incident polarization angle $\phi=0$ is shown in Fig.~\ref{angles_TMD}(a). 
Notably, the calculated values of $\theta_K$ and $\zeta_K$ are nearly identical with SOC turned on or off, demonstrating that SOC plays a negligible role in the obtained polarization rotation response. 
The sign reversals of $\theta_{\rm K}$ are mainly due to those of the quantum-metric nematicity $\bar{g}_{x^2-y^2}$ and $\bar{g}_{xy}$, as discussed in details in End Matter.
We then fix the photon energy at $\hbar\omega = 3.4$ eV, corresponding to a pronounced peak in both $\theta_K$ and $\zeta_K$, and examine the signal as a function of the initial polarization direction $\phi$ [Fig.~\ref{angles_TMD}(b)]. The results remain qualitatively unchanged irrespective of SOC. These findings confirm that a QMNKR can indeed be realized in realistic transition metal dichalcogenides materials preserving $\mathcal{T}$ symmetry, and that its signature is largely independent of SOC.
Importantly, the $\theta_K$ still shows a $\text{sin}(2\phi+\phi_0)$-like behavior.
Furthermore, the magnitude of the obtained $\theta_K$ and $\zeta_K$ lies well within experimentally measurable values~\cite{Huang2017, Xu2022, Azzam1989}. 
Thus, the QMNKR serves as a promising probe of quantum-metric nematicity in realistic materials.

\emph{Discussions.}---The results obtained in the above demonstrate that the QMNKR directly occurs as the result of nontrivial quantum geometry, particularly capturing the quantum-metric nematicity of the materials.
It has also been demonstrated that the polarization rotation and ellipticity can arise even in nonmagnetic systems with $\mathcal{T}$ symmetry and in the absence of SOC, as the consequence of the quantum-metric nematicity.
Besides probing the quantum-metric nematicity,
an important application of the QMNKR is to detect the nematic transition of the system and visualization of nematic domains~\cite{Xu2022}, even in atomically thin samples.
Thus, the QMNKR offers a contact-free probe of the nematic order and the quantum-metric nematicity brought with it.

 It should be noted that angle-dependent polarization-rotation effects are often described phenomenologically in terms of birefringence or anisotropic optical response~\cite{Xu2022}.
However, we point out that there is a quantum-metric-nematicity-induced polarization rotation even in atomically thin materials, even with $\mathcal{T}$-symmetry, beyond such phenomenological birefringence explanation.


We thank Qian Niu, Wenyu He, Yingming Xie, Xunjiang Luo, Xilin Feng and Jinxin Hu for valuable discussions.
A.D. thanks Cheng-Cheng Liu and Shingo Yonezawa for helpful email correspondence. AD is supported by JSPS KAKENHI (Grant No. JP21K13880, JP23K17353, JP24H01662, JP23KK0248).
K. T. L. acknowledges the support of the Ministry of Science and Technology, China, The New Cornerstone Foundation, and the Hong Kong Research Grants Council through Grants No. MOST23SC01-A, No. RFS2021-6S03, No. C6053-23G, No. AoE/P-701/20, AoE/P-604/25R, No. 16309223, No. 16311424 and No. 16300325.
 
\bibliography{kerr}

\newpage
\appendix
\begin{center}
\textbf{End Matter}    
\end{center}

\section{Results of the optical conductivity and quantum metric nematicity in strained MoS$_2$}

\begin{figure*}
	\includegraphics[width=17cm,angle=0]{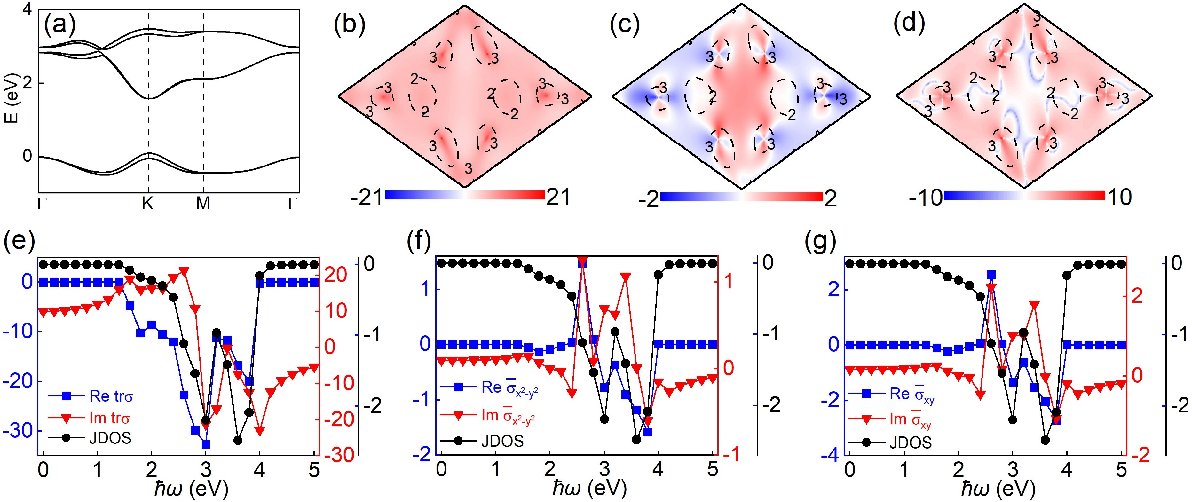}
	\centering
	\caption{The quantum metric nematicity and optical conductivity (in units of $e^2/h$) in strained MoS$_2$. (a) The band structure along the high symmetry lines. The distribution of the logarithm of the quantum metric qualities (b) $\text{tr} g$, (c) $\bar{g}_{x^2-y^2}$, and (d) $\bar{g}_{xy}$ in the first Brillouin zone. The optical conductivity (e) $\text{tr} \sigma$, (f) $\bar{\sigma}_{x^2-y^2}$, (g) $\bar{\sigma}_{xy}$ and corresponding JDOS.}
	\label{TMD1}
\end{figure*}

We use the model for monolayer MoS$_2$ from previous study with parameters determined by first-principle calculations\cite{SM}. 
The band structure has a gap at $K$ point with spin splitting due to the SOC as shown in Fig.~\ref{TMD1}(a). Due to the highly localized distribution of the quantum metric across the Brillouin zone, we plot the logarithm of the quantum metric to resolve its spatial variation more clearly. The anisotropy of the system manifests in finite $\bar{g}_{x^2-y^2}$ and $\bar{g}_{xy}$ displayed in Fig.~\ref{TMD1}(c) and (d), respectively. Correspondingly, the optical conductivity $\text{tr} \sigma$, $\bar{\sigma}_{x^2-y^2}$, and $\bar{\sigma}_{xy}$ are shown in Fig.~\ref{TMD1}(e)-(g). Together with the energy weight factor and the JDOS, quantum metric nematicity gives both the peaks and the sign change of the optical conductivity. In addition, consistent with the results from the minimal model discussed earlier, the optical conductivity is negligibly small for photon energies smaller than the band gap.

\end{document}